\documentclass[footinbib,twocolumn,aps,prb,amsmath,amssymb,superscriptaddress,longbibliography,aip,floatfix]{revtex4-2}

\usepackage{lipsum}
\usepackage{verbatim}
\usepackage[normalem]{ulem}
\usepackage[colorlinks=true,linkcolor=blue,citecolor=blue]{hyperref}
\usepackage[utf8]{inputenc}
\usepackage[T1]{fontenc}

\usepackage{bm}
\usepackage{array}
\usepackage{mathtools}
\usepackage[mathscr]{euscript}
\usepackage{mathrsfs}

\usepackage{physics}
\usepackage{algorithmic}
\usepackage{algorithm}
\usepackage{graphicx}
\usepackage{float}
\usepackage{tikz}
\usepackage{pgfplots}
\usepackage[caption=false]{subfig}
\usepackage{epstopdf}

\DeclareMathOperator*{\argmax}{argmax}

\newcommand{\blue}[1]{{\color{blue}{#1}}}

\begin{document}

\title{Simulation of interaction-induced chiral topological dynamics on a \\ digital quantum computer}

\author{Jin Ming Koh}
\affiliation{Division of Physics, Mathematics and Astronomy, Caltech, Pasadena, California 91125, USA}
\author{Tommy Tai}
\affiliation{Cavendish Laboratory, University of Cambridge, JJ Thomson Ave, Cambridge CB3 0HE, UK}
\affiliation{Department of Physics, National University of Singapore, Singapore 117542}
\author{Ching Hua Lee}
\email{phylch@nus.edu.sg}
\affiliation{Department of Physics, National University of Singapore, Singapore 117542}
\date{\today}
\begin{abstract}
Chiral edge states are highly sought-after as paradigmatic topological states relevant to both quantum information processing and dissipationless electron transport. Using superconducting transmon-based quantum computers, we demonstrate chiral topological propagation that is induced by suitably designed interactions, instead of flux or spin-orbit coupling. Also different from conventional 2D realizations, our effective Chern lattice is implemented on a much smaller equivalent 1D spin chain, with sequences of entangling gates encapsulating the required time-reversal breaking. By taking advantage of the quantum nature of the platform, we circumvented difficulties from the limited qubit number and gate fidelity in present-day noisy intermediate-scale quantum (NISQ)-era quantum computers, paving the way for the quantum simulation of more sophisticated topological states on very rapidly developing quantum hardware.  
\end{abstract}

\maketitle

\textit{Introduction.}---The discovery of the integer quantum Hall effect in 2D electron gases revolutionized condensed matter physics~\cite{halperin1982quantized,macdonald1984quantized}. To circumvent the requirement of a strong external magnetic field, tight-binding lattices with intrinsic time-reversal symmetry (TRS) breaking have been devised, inspired by Haldane's seminal Chern insulator~\cite{haldane1988model}. Such lattice-based chiral topological phenomenon is known as the `\emph{quantum anomalous Hall effect}' and has been demonstrated in ferromagnetic topological insulators~\cite{PhysRevLett.101.146802,yu2010quantized,chang2013experimental,zhao2020tuning}, magic-angle twisted bilayer graphene~\cite{geisenhof2021quantum, tschirhart2021imaging, serlin2020intrinsic, sharpe2019emergent}, and Moir\'{e} heterostructures~\cite{chen2020tunable,polshyn2020electrical,zhang2019nearly}. Such is their academic and potential technological impact that topological boundary states have also been realized in photonic platforms~\cite{PhysRevLett.100.013904,fu2010robust,wang2009observation,lin2017line,PhysRevX.5.031011,lu2014topological,khanikaev2017two,ozawa2019topological,smirnova2020nonlinear}, polariton~\cite{klembt2018exciton,kartashov2019two,mandal2020nonreciprocal,su2021optical}, mechanical gyrotopic~\cite{kane2014topological,nash2015topological,PhysRevB.97.085110} and acoustic~\cite{khanikaev2015topologically,PhysRevLett.115.104302,susstrunk2015observation,he2016acoustic,PhysRevX.5.031011,Guo2021} systems, as well as topolectrical circuits~\cite{hofmann2019chiral,lee2018topolectrical,PhysRevX.5.021031, PhysRevB.99.161114,albert2015topological,cai2019observation,ezawa2019non,zhang2019topological,olekhno2020topological,lee2020imaging,ezawa2021topological,stegmaier2021topological,zou2021observation,shang2022experimental,zhang2022anomalous,yang2022experimental, hohmann2022observation,liu2022observation}. 

In the above, the broken TRS required for chiral topological propagation have been introduced through %magnetic dopants in ferromagnetic topological insulators, and via orbital magnetism in bilayer graphene and Moir\'{e} heterostructures. 
magnetic dopants, orbital magnetism or non-reciprocal media. Tantalizing alternative routes, however, exist when the chiral lattice is physically implemented on universal quantum simulators, as in quantum circuits and quantum computers~\cite{PRXQuantum.2.017003, choo2018measurement, smith2019crossing, azses2020identification, mei2020digital, koh2021stabilizing, tan2021realizing}. In particular, such fully quantum platforms have the innate propensity to realize the nontrivial TRS-breaking topology via novel many-body effects and interactions. In this work, we demonstrate how chiral topological dynamics can be induced through engineered interactions, rather than conventional single-body mechanisms such as spin-orbit coupling.

As intrinsically quantum platforms, quantum computers are in principle able to realize any quantum phenomenon, including strong interaction effects beyond the reach of classical topological metamaterials. Even in the current noisy intermediate-state quantum (NISQ) era, digital quantum computers have shown incredible promise~\cite{zhong2021quantum, satzinger2021realizing, google2020hartree, chiesa2019quantum, arute2019quantum,arute2019quantum} due to their versatility, complementing alternative experimental platforms such as ultracold polar molecules~\cite{moses2017new,PhysRevX.9.021039} and Rydberg atoms~\cite{samajdar2021quantum, de2019observation,ebadi2021quantum}. Programmable with universal quantum gate sets that can simulate generic unitaries, quantum computers are ideally poised as platforms for observing various condensed-matter and topological phenomena, particularly those that require esoteric or many-body coupling configurations. Existing bottlenecks, such as low gate fidelity, decoherence, limited qubit connectivity and limited number of qubits~\cite{preskill2018quantum}, may be significantly alleviated by error mitigation and circuit optimization techniques~\cite{kandala2019error, kandala2017hardware, temme2017error, cerezo2021variational, heim2020quantum}. 

To date, simulating two-dimensional systems such as Chern lattices has been inherently challenging given the aforementioned hardware constraints. Physical simulations on digital quantum computers have largely been restricted to 1D spin or fermionic chains and small 2D systems \cite{kirmani2022probing, rahmani2020creating, smith2019simulating, zhu2021probing, cervera2018exact, paulson2021simulating,mi2022time}. Indeed, due to the numerous lattice sites ($\gtrsim 10^2$) necessary for cleanly observing chiral mode propagation, these paradigmatic topological states have been missing from quantum computer demonstrations.

\begin{figure}
\includegraphics[width = 0.98\linewidth]{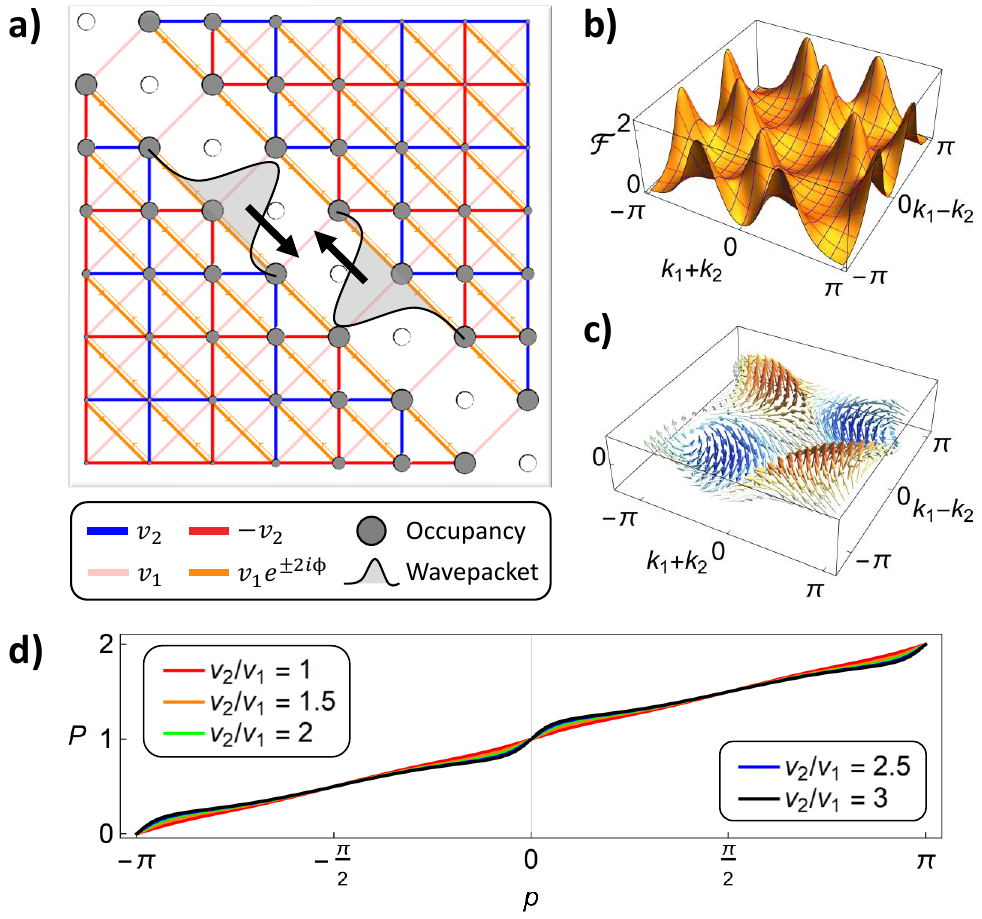}
\caption{\textbf{Chiral topological propagation from interactions.} \textbf{(a)} Chern topology arises from 
a checkerboard lattice with non-reciprocal couplings carrying effective flux (thick and thin orange arrows). Implemented as a 1D interacting chain, vertical/horizontal and diagonal couplings become single-boson (red, blue) and two-boson hoppings (pink, orange) respectively. An on-site repulsive interaction is imposed to yield a virtual boundary along the lattice diagonal, along which chiral propagation occurs in the form of oppositely moving two-particle wavepackets (black arrows). \textbf{(b)} The Berry curvature $\mathcal{F}$ and \textbf{(c)}  its corresponding $\vb{d}(\vb{k})$-vector skyrmionic texture of our model for $v_1 = v_2 = 1$ and $\phi =\pi/6$. The total number of chiral edge modes is the Chern number $\smash{\int\mathcal{F} \,  \dd[2]{\vb{k}} / 2\pi =2}$. \textbf{(d)} The Wannier polarization representing the center of mass translation of a maximally-localized wavepacket as we thread a flux $\smash{k_+ \rightarrow k_+ + A_y}$. It is pumped $C=2$ upon a complete cycle, and moves with almost perfect uniformity for most values of the ratios $v_2/v_1$.}
\label{fig:main-fig-1}
\end{figure}

In this \textit{Letter}, we propagate a chiral topological mode on a quantum computer, demonstrating a versatile quantum simulation setup that complements existing experimental realizations of Chern topology. By designing logical two-body interactions, the Chern lattice is mapped exactly onto a 1D chain with two interacting excitations~\footnote{This work serves to 2D 1-body Chern phenomena by mapping to a 1D 2-body problem, while a few existing theoretical and experimental works aims to understanding other 1D 2-body systems by studying 2 1-body setups~\cite{longhi2013tamm,mukherjee2016observation,gorlach2018analogue,olekhno2020topological}.}, which is well within the capabilities of NISQ quantum computers. Our remapping relied on the fact that Chern topological states are fundamentally single-body phenomena, and can thus be ``compressed'' into a smaller physical system already hosting an exponentially large quantum many-body Hilbert space. We employed IBM transmon-based superconducting quantum computers, which has been utilized in various applications~\cite{havlicek2019supervised, koh2022experimental}, including quantum chemistry~\cite{mcCaskey2019quantum, smart2020efficient} and spin lattice problems~\cite{zhukov2018algorithmic, smith2019simulating}. Compared to existing quantum computer simulations of other topological systems~\cite{choo2018measurement, smith2019crossing, azses2020identification}, ours is performed in physical, not synthetic, space with a significantly more complex Hamiltonian, allowing for the observation of topological boundary chiral dynamics in an intrinsically 2D setting.

\textit{Chern lattice as an interacting model.}---A Chern lattice is a system with an integer $C$ number of robust chiral boundary modes, where $C$ is the Chern topological invariant~\cite{TKNN}. This invariant is defined as the index of the mapping $T^2\rightarrow \Gamma$, where $\Gamma$ is the state space where eigenstates live in. For 2-band models, $\Gamma$ takes the familiar form of the Bloch sphere, and $C$ measures how many times the torus $T^2$ wraps around the sphere. 

Although the torus is commonly taken to represent 2D lattice momentum space, it can also refer to the joint configuration momentum space of two particles, whose interactions correspond to effective hoppings that shape the mapping $T^2\rightarrow \Gamma$. This alternative representation takes advantage of the much larger many-body Hilbert space intrinsically accessible on a quantum computer: unlike in classical systems, just $N$ qubits can index up to $2^N$ possible configurations, thereby easily accommodating the $\order*{N^2}$-dimensional basis of $T^2$. As such, by rewriting certain hoppings in terms of interaction terms, we can achieve drastically more compact simulations of higher-dimensional lattices. Such a mapping makes Chern boundary modes, which usually require $\order*{10^2}$ lattice sites~\footnote{Assuming a minimum of 5 to 7 unit cells in each 2D direction, with two sites per unit cell.} to properly resolve, readily observable on current-generation quantum computers with $\order*{10^1}$ high-quality interconnected qubits.

Besides allowing for the compact simulation of physical lattices, a digital quantum computer realization is greatly versatile in accommodating couplings of differing ranges and types. For concreteness, we discuss in the context of the topological Hamiltonian that we implemented; but this versatility applies to lattice Hamiltonians in general. We used a modified version of the checkerboard lattice model well-known for flat band properties and strongly-correlated states~\cite{sun2011nearly,lee2014lattice,regnault2011fractional}, $\mathcal{H}=\sum_{\vb{k}} \mathcal{H}(\vb{k}) c^\dagger_{\vb{k}} c_{\vb{k}}$, where $\mathcal{H}(\vb{k})=\vb{d}(\vb{k})\cdot \vb*{\sigma}=$
\begin{equation}
\begin{bmatrix}-4v_2\sin k_-\sin k_+&2v_1(\cos k_-+\cos k_+e^{2i\phi})\\2v_1(\cos k_-+\cos k_+e^{-2i\phi})&4v_2\sin k_-\sin k_+\\\end{bmatrix},
\label{paradigmaticChern}
\end{equation}
for $k_\pm=k_1\pm k_2$ written in its original 2D momentum space. Above, $\vb{d}(\vb{k})$ denotes the Bloch vector (see Supplement \blue{S1}) and $\vb*{\sigma}$ are Pauli matrices. The lattice-space representation of our Hamiltonian, obtained by a Fourier transform of Eq.~(\ref{paradigmaticChern}), is exactly depicted in Fig.~\ref{fig:main-fig-1}\blue{a}, consisting of two square lattices interlocked in a checkerboard fashion. This lattice contains inter-sublattice couplings $v_1$ in the directions $\Delta\vb{r} = (1,\pm 1)$ and $(-1,\pm 1)$. Vertical and horizontal $v_2$ hoppings take alternate signs in adjacent checkers (Fig.~\ref{fig:main-fig-1}\blue{a}). Importantly, the requisite TRS breaking for Chern topology enters through the alternating phase factors $e^{\pm 2i\phi}$ along the diagonal $\Delta \vb{r} \propto (1, 1)$, in the couplings between the upper/lower and lower/upper sublattices. While realizing these phase factors can be challenging in conventional topological media~\cite{Guo2021, wirth2011evidence, landig2016quantum, PhysRevA.91.033604, PhysRevLett.106.015302, PhysRevLett.107.140402}, they are readily achievable on a digital quantum computer.

\begin{comment}
Hamiltonian $H_{\text{2D}}$:
\begin{equation}\begin{split}
H_{\text{1D}}&=v_1\sum_{\lambda=\pm1,i}(\mu_i\nu_i^\dag+\mu_i^\dag\nu_i^\dag e^{\lambda2i\phi})\nonumber\\&+v_2\sum_{i}(-1)^i(\mu_{i+2}^\dag\mu_i-\nu_{i+2}^\dag\nu_i)+\text{h.c.}\end{split}
\end{equation}
\end{comment}

As the next step, we rewrite these hoppings via $c^\dagger_{x_1, x_2} \rightarrow \mu^\dagger_{x_1} \nu^\dagger_{x_2}$, where $\mu^\dagger$ and $\nu^\dagger$ create hardcore bosons of two different species in 1D, such that the $(1+1)$-body sector of this new system $\mathcal{H}_\text{1D}$ corresponds to the original 2D Hamiltonian $\mathcal{H}$ with
\begin{equation}\begin{split}
\mathcal{H}_\text{1D} 
    &= v_1\sum_{x_1, x_2 \, \text{even}} 
        \left( \mu^\dagger_{x_1+1} e^{2i\phi} + \mu^\dagger_{x_1-1} \right) 
        \nu^\dagger_{x_2+1} \mu_{x_1} \nu_{x_2} \\
    &+ v_1 \sum_{x_1, x_2 \, \text{odd}}
        \left( \mu^\dagger_{x_1+1} e^{-2i\phi} + \mu^\dagger_{x_1-1} \right)
        \nu^\dagger_{x_2+1} \mu_{x_1} \nu_{x_2} \\
    &+ v_2 \, \sum_{x} (-1)^x (\mu^\dagger_{x+2}\mu_{x} - \nu^\dagger_{x+2}\nu_{x}) + \text{h.c.}.
\end{split}
\label{eq:H1D-topo}
\end{equation}
In this formulation, the $v_2$ hoppings, which were originally vertical or horizontal, remain single-body, albeit over next nearest neighbors. But, the diagonal $v_1$ hoppings become simultaneous two-body hopping interactions, some containing phase rotations $e^{\pm 2i\phi}$. Further, the impenetrability of the hardcore $\mu,\nu$-bosons (see Supplement \blue{S1}) enforces a virtual boundary that prohibits double site occupancy along $x_1=x_2$, \textit{i.e.}, the diagonal line perpendicular to $k_-$. The interference of the phase rotations collude to give rise to chiral topological transport along this boundary. This is apparent upon referring to the polarization 
\begin{equation}
P(p_+)=\frac1{4\pi}\int_{-\pi}^{p_+}\int_{-\pi}^{\pi}\mathcal{F}(k_+,k_-) \, \dd{k_-} \dd{k_+},
\end{equation}
which changes by $C=2$ sites upon a complete cycle of flux threading (Fig.~\ref{fig:main-fig-1}\blue{d}), implying non-trivial topological pumping. Here, $\mathcal{F}=-i\langle\partial_{x_1}\psi|\partial_{x_2} \psi\rangle + \text{h.c.}$ is the Berry curvature \footnote{$\mathcal{F}$ is a measure of the topological charge, giving a Chern number of $C=\int_{-\pi}^{\pi}\int_{-\pi}^{\pi} \mathcal{F} \, \dd{k_1} \dd{k_2} /2 \pi = 2$, as expected.}, computed from the Jacobian determinant of the $\vb{d}(\vb{k})$ mapping (Figs.~\ref{fig:main-fig-1}\blue{c}, \ref{fig:main-fig-1}\blue{d}). Physically, $P$ reflects the spectral flow propagation of Wannier centers~\cite{PhysRevB.84.075119,resta1998quantum,lee2015free,nelson2021multicellularity} upon flux threading, and for this model, can be optimized to be as uniform as possible to facilitate uniform chiral boundary propagation (Supplement \blue{S1}).  In terms of the $\mu,\nu$-bosons, this pumping is manifested as the motion of correlated particle pairs occupying adjacent sites, which is robustly protected by the flux asymmetry caused by the dissimilar effects of the phases $\phi$ on the left and the right of each boson. Along the virtual boundary, where the two bosons are next to each other, this asymmetry leads to a correlated pumping of their center-of-mass, \textit{i.e.} chiral propagation.

\begin{figure}
    \includegraphics[width=0.97\linewidth]{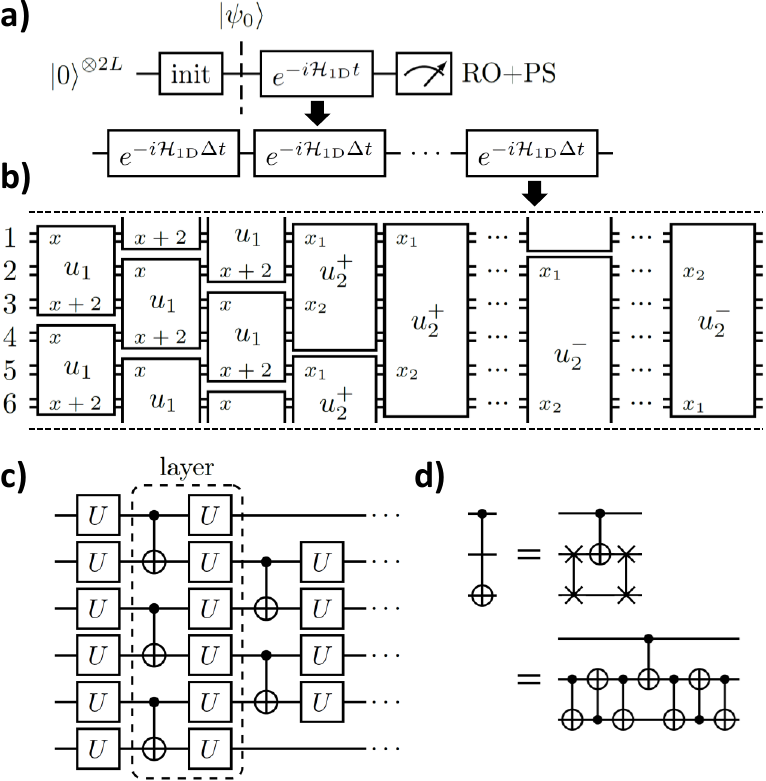}
    \caption{\textbf{Quantum circuit implementation schematics.} \textbf{(a)} Schematic of quantum circuit for time-evolving an initial state $\ket{\psi_0}$ according to our Hamiltonian $\mathcal{H}_\text{1D}$, with error mitigation (RO$+$PS) applied on the final measurements.  \textbf{(b)} In-principle first-order trotterization of the $e^{-i\mathcal{H}_\text{1D} t}$ propagator, and the breakdown of a trotter step in terms of unitaries $u_1$ and $u_2$, as further elaborated in Supplement \blue{S2}. Pairs of qubits (drawn as double lines) represent each site of the logical Chern lattice. \textbf{(c)} Circuit ansatz for recompilation, comprising layers of entangling CXs and single-qubit rotations, stacked in brickwork pattern. \textbf{(d)} Non-nearest neighbour CX gates can be implemented using SWAPs, which themselves decompose into nearest-neighbour CXs.}
    \label{fig:main-fig-2}
\end{figure}

\begin{figure*}
    \centering
    \includegraphics[width=0.9\linewidth]{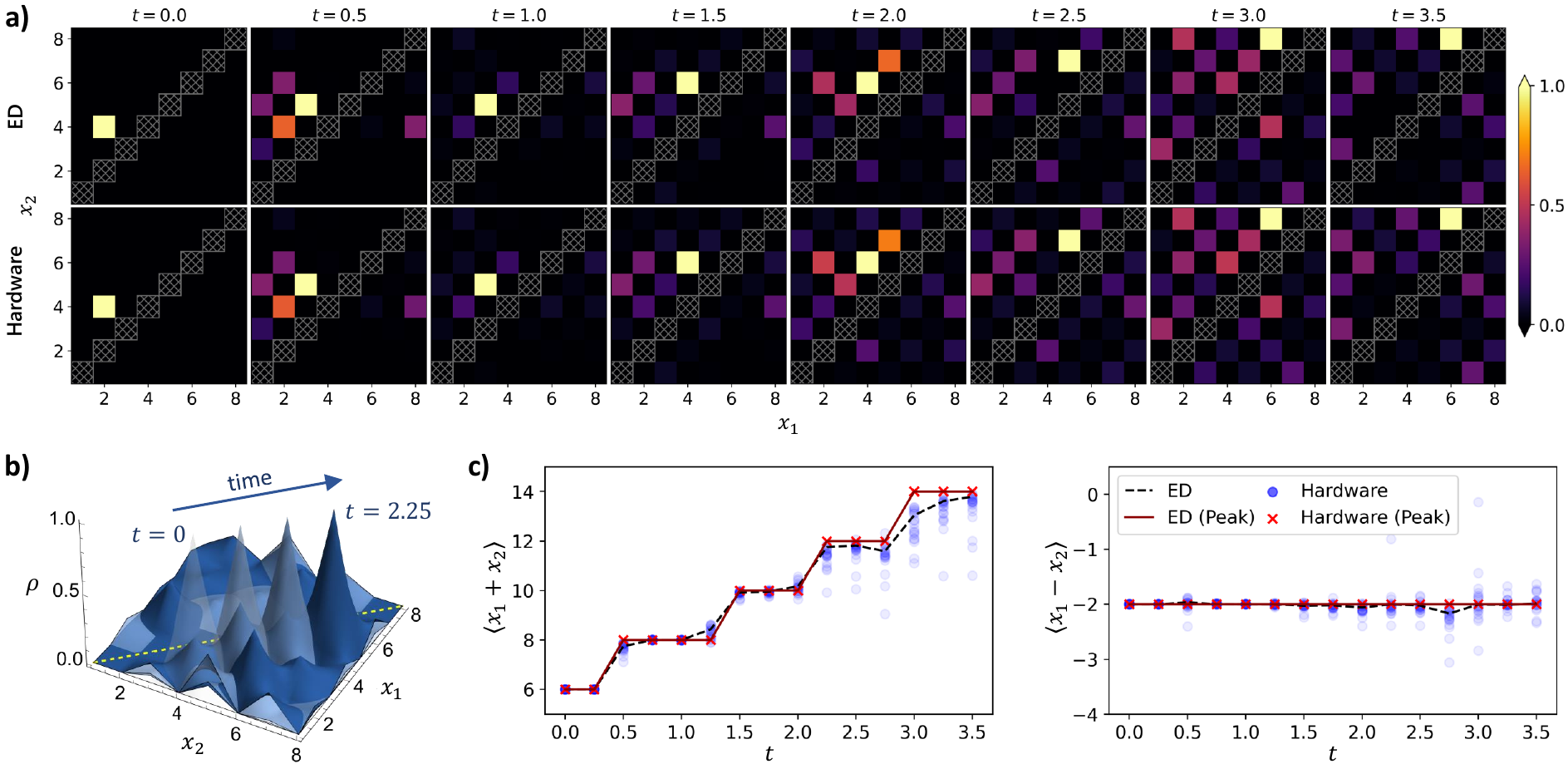}
    \caption{\textbf{Demonstration of chiral topological propagation on a quantum computer}. \textbf{(a)} Evolution of occupancy densities $\smash{\rho(x_1, x_2) = \expval{\mu^\dagger_{x_1} \mu_{x_1} \nu^\dagger_{x_2} \nu_{x_2}}}$ as time progresses, with good agreement between exact diagonalization (top row) and hardware data (bottom row), for an initial wavepacket at $(x_1, x_2) = (2, 4)$. We have normalized the peak of $\rho(x_1, x_2)$ for visual clarity. Hatch-shaded squares indicate the $x_1 = x_2$ virtual boundary. \textbf{(b)} 3D visualization of the same chiral propagation, which clearly reveals the peak translating parallel to the $x_1 = x_2$ virtual boundary. The sequence of snapshots of measured $\rho(x_1, x_2)$ at $t=0, 0.75, 1.50$ and $2.25$ are superimposed (lightest to darkest). \textbf{(c)} (Left) Monotonic evolution of the center-of-mass $\langle x_1+x_2\rangle$, with localized peak positions (crosses) shifting together with the density profile (blue) and centroid $\smash{\expval{x_1 + x_2} = \sum_{x_1, x_2} (x_1 + x_2) \rho(x_1, x_2)^4}$ (dashed). Exact diagonalization (ED) results agree closely with data from the quantum hardware. (Right) The state remains close to the diagonal virtual boundary, as indicated by constant $\expval{x_1 - x_2} = 2$ over time. Parameters are $v_1 = v_2 = 1$ and $\phi = \pi/6$.} 
    \label{fig:main-fig-3}
\end{figure*}

\textit{Realization on a quantum computer.}---To simulate chiral propagation, we implement $\mathcal{H}_\text{1D}$ on IBM transmon-based quantum computers~\cite{garcia2020ibm, cross2018ibm, Qiskit}. The quantum nature of this platform allows many-body systems to be directly simulated; in the present context of $\mathcal{H}_\text{1D}$, we represent each unit cell of the Chern lattice model with $2$ qubits, for a total of $16$ qubits representing an $L = 8$ lattice. This is a considerable reduction from the $8^2 = 64$ qubits otherwise needed, without the dimensional reduction into a 1D system. We use 27-qubit quantum devices in our simulation runs (see Supplement \blue{S2}).

We map the $\{\ket{00}, \ket{01}, \ket{10}\}$ computational basis states of each pair of qubits to unoccupied, occupied by a $\mu$-boson, and occupied by a $\nu$-boson states of the model. The hardcore bosonic constraint is enforced by an appropriate representation of $\mu, \nu$ operators satisfying the canonical mixed-commutation relations. An initial state $\ket{\psi_0}$ evolves over time as $\ket{\psi(t)} = U(t) \ket{\psi_0}$, with propagator $U(t) = e^{-i \mathcal{H}_\text{1D} t}$. To study state dynamics on the quantum computer, it is necessary to implement $U(t)$ as a quantum circuit; a standard method is to decompose $\smash{\mathcal{H}_\text{1D} = \sum_{\vb*{\gamma}} A_{\vb*{\gamma}} \sigma^{\vb*{\gamma}}}$ in the spin-$1/2$ basis, for generically non-commuting Pauli strings $\sigma^{\vb*{\gamma}}$ and $A_{\vb*{\gamma}} \in \mathbb{R}$, and employ trotterization~\cite{georgescu2014quantum, ortiz2001quantum}. In the first-order scheme, we split $\smash{e^{-i \mathcal{H}_\text{1D} t} = \left(e^{-i \mathcal{H}_\text{1D} \Delta t}\right)^n}$ into $n$ steps, each of which is approximated as $\smash{\prod_{\vb*{\gamma}} e^{-i A_{\vb*{\gamma}} \Delta t}}$. Here, it is convenient to separate $\mathcal{H}_\text{1D} = h_1 + h_2$, such that $h_1$ contains the single-boson hoppings, and $h_2$ the two-boson interactions. Then the trotterized circuit can be constructed from unitaries $u_1$ and $u_2$, respectively implementing evolution by $h_1$ and $h_2$, as shown in Fig.~\ref{fig:main-fig-2}\blue{b}. See Supplement \blue{S2} for further details.

Although the physical couplings between the hardware qubits are only between nearest neighbors, long-ranged entangling gates corresponding to distant couplings in $u_1$ and $u_2$ can be effected with the application of SWAP gates. For instance, at the basic level, non-nearest neighbour CX gates can be effected by chaining CXs on adjacent qubits (Fig.~\ref{fig:main-fig-2}\blue{d}). The single-particle hoppings and two-particle quartic interactions are treated on equal footing in the quantum circuit implementation, and the simulation is digital in nature, compared to the aforementioned analog classical platforms~\cite{PhysRevLett.100.013904,fu2010robust,wang2009observation,PhysRevX.5.031011,kane2014topological,nash2015topological,PhysRevB.97.085110,khanikaev2015topologically,PhysRevLett.115.104302,susstrunk2015observation,he2016acoustic,PhysRevX.5.031011,Guo2021,hofmann2019chiral, PhysRevX.5.021031, lee2018topolectrical, imhof2018topolectrical, PhysRevB.99.161114, wang2019topologically, lee2020imaging}.

We note, however, that in this form, the trotterized circuits for $\mathcal{H}_\text{1D}$ are infeasibly deep for current NISQ-era devices. Present benchmarks indicate $\order{10^2}$ CX layers are achievable with state-of-the-art techniques and hardware, likely yet smaller for quantitatively accurate simulations; the numerous trotter steps required for acceptable truncation error easily exceeds this limit. We thus employ an implementation strategy known as circuit recompilation~\cite{sun2021quantum, koh2021stabilizing, khatri2019quantum, heya2018variational} to compress circuit depth. A circuit ansatz (Fig.~\ref{fig:main-fig-2}\blue{c}) comprising layers of single-qubit rotations and CX entangling gates, laid in a brickwork pattern, is iteratively optimized through tensor network-aided quantum simulation~\cite{gray2018quimb} to approach the intended unitary. Specifically, we collect the rotation angles $\vb*{\vartheta} = (\vb*{\theta}, \vb*{\phi}, \vb*{\lambda})$, and numerically treat the optimization problem $\smash{\argmax_{\vb*{\vartheta}} \abs{\smash{\mel{\psi_0}{V_{\vb*{\vartheta}}^\dagger U}{\psi_0}}}^2}$ for circuit unitary $V_{\vb*{\vartheta}}$. Additionally, we note $\mathcal{H}_\text{1D}$ is number-conserving in both the $\mu$ and $\nu$ species, and also conserves the parity of $x_1 + x_2$. As our focus is in states $\ket{\psi_0}$ with definite $\mu, \nu$-particle numbers and $x_1+x_2$ parity (corresponding to the single-particle sector on the checkerboard lattice), we need only perform calculations over the symmetry-restricted sector, significantly alleviating costs. This sector-specific recompilation technique hugely reduces circuit depth (to $\leq 12$ CX layers), and is critical in our realization of the $\mathcal{H}_\text{1D}$ model on present hardware. See Supplement \blue{S2} for technical details.

To suppress the effects of hardware noise, we employ readout error mitigation~\cite{kandala2019error, kandala2017hardware, temme2017error}, post-selection~\cite{mcardle2019error, smith2019simulating}, and averaging across qubit chains and machines. In particular, we run calibration circuits alongside each experiment, and approximately correct measurement bit-flip errors via linear inversion; to feasibly accommodate $16$ qubits, we use an approximate tensored scheme \cite{koh2021stabilizing}. Finally, results that violate the $\mu, \nu$-particle number and $x_1+x_2$ parity symmetries of $\mathcal{H}_\text{1D}$ are non-physical, enabling a post-selection policy at no additional circuit depth nor measurement costs.

\textit{Measured results.}---We directly measured interaction-induced topological chiral propagation (Fig.~\ref{fig:main-fig-3}) along a diagonal virtual boundary $x_1 = x_2$ in configuration space. We simulate the time-evolution of an initial state localized at position $(x_1, x_2) = (2, 4)$ adjacent to the diagonal; an analogous set of results for $(x_1, x_2) = (7, 5)$ on the opposite side of the diagonal is given in Supplement \blue{S3}. In both, the site-resolved occupancy densities $\rho(x_1, x_2)$ measured on hardware very closely match exact diagonalization results. The chiral nature of the boundary-localized propagating modes are more saliently presented in  (Figs.~\ref{fig:main-fig-3}\blue{b--d}), where unidirectional propagation along the diagonal, in opposite directions for the two initial states, is clearly observed, as indeed expected. The unidirectionality of transport is apparent from the movement of the wavepacket peak, which closely track $\expval{x_1 + x_2}$ and monotonously increases; localization along the virtual boundary is maintained throughout transport, as verified by an almost constant $\expval{x_1 - x_2}$. 

While the presently studied model is interesting as it hosts chiral modes under periodic boundary conditions without physical edges, we demonstrate chiral propagation also under more traditional open boundary conditions (see Supplement \blue{S3}). On such a lattice, boundary-localized wavepackets unidirectionally propagate along the virtual boundary of the diagonal, bend at the corner and travel along the physical edges, and loop back onto the diagonal. In both cases, we successfully demonstrate quantum anomalous Hall-type topological transport on a fully programmable quantum platform, despite the inherent limitations of current-day NISQ-era devices.

\textit{Discussion.}---In this work, we successfully demonstrated chiral propagating modes on a quantum computer. While observing such Chern propagation typically requires implementing 2D quantum anomalous Hall lattices, we instead employed a checkerboard model that maps into a 1D interacting chain with Pauli-like repulsive interactions. Through a combination of remapping the lattice to a spin chain and the use of appropriately-designed many-body interactions~\cite{olekhno2020topological,lee2021many}, we were able to circumvent circuit breadth limitations on present-day noisy intermediate-scale quantum (NISQ)-era quantum computers.

Indeed, quantum computers provide a versatile platform for investigating new condensed-matter phenomena, including those involving long-ranged or exotic many-body couplings. With a Hilbert space that harbors exponentially many degrees of freedom, they can accommodate various phenomena such as many-body scars, quantum phase transitions, and time crystals, even when the requisite interactions are challenging to realize on other platforms. While the full advantage of digital quantum simulation can only be realized with quantum error correction and fault-tolerant quantum computing, recent progress, including the present study, demonstrate already encouraging capabilities on rapidly advancing NISQ-era hardware. As a final remark, we point out that while we have focused on the quantum simulation of a 2D Chern insulator, our mapping method is general to arbitrary $d$-dimensional, $n$-band one-body systems (see Supplement \blue{S1}), and is readily applicable to study more sophisticated topological phenomena~\cite{lee2015geometric,chen2019direct,gu2016holographic,tuloup2020nonlinearity,peterson2020fractional,lee2018electromagnetic,yang2020photonic,kirmani2022probing, rahmani2020creating}\footnote{For example, the paradigmatic Hofstadter model, which contains hopping with more general site-dependent phase factors on a rectangular lattice in comparison to the presently considered checkerboard model. We provide a discussion of its mapping and implementation in Supplement \blue{S1}, which also referenced Refs.~\cite{PhysRevB.14.2239,PhysRevLett.45.494,PhysRevB.23.5632, PhysRevB.27.6083,PhysRevX.1.021014,PhysRevB.2.723,ginibre1969existence,trotter1959product,montvay1997quantum,malouf2002comparison,vidal2004efficient,hatano2005finding,andrew2007scalable,moll2018quantum,alvarez2018quantum,behera2019designing,sieberer2019digital,heyl2019quantum,cross2019validating,motta2020determining,jones2020quantum,qiskit2020textbook}.}.

\textit{Acknowledgements.}---J.M.K. and T.T. thank Wei En Ng and Yong Han Phee of the National University of Singapore for discussions on the quantum simulation implementation. This work is supported by the Singapore National Research Foundation grant award no. NRF2021-QEP2-02-P09. The authors acknowledge the use of IBM Quantum services for this work. The views expressed are those of the authors, and do not reflect the official policy or position of IBM or the IBM Quantum team.

\bibliography{ref-topo,ref-qc}

\end{document}